\documentstyle[prl,aps,multicol,epsfig]{revtex} 

\def\etal{\emph{et al.}\ }

\def\um{$\mu$m}

\def\uohmcm{$\mu\Omega\cdot$cm}

\begin{document}
\draft
\title{Mesoscopic Ferromagnet/Superconductor Junctions and the 
Proximity Effect}
\author{J.\ Aumentado and V.\ Chandrasekhar}
\address{Department of Physics and Astronomy, Northwestern University, Evanston, IL
60208} 
\maketitle

\begin{abstract} 
We have measured the electrical transport of submicron ferromagnets (Ni) in contact 
with a mesoscopic superconductor (Al) for a range of interface 
resistances. In the geometry measured, the 
interface and the ferromagnet are measured separately. The ferromagnet 
itself shows no appreciable superconducting proximity effect, but the 
ferromagnet/superconductor interface exhibits strong 
temperature, field and current bias dependences.  
These effects are dependent on the local magnetic field distribution near 
the interface arising from the ferromagnet. 
We find that the temperature dependences may be 
fit to a modified version of the Blonder-Tinkham-Klapwijk 
theory for normal-superconductor transport.
\end{abstract}

\begin{multicols}{2}
There has been much interest recently about the 
possibility of observing the superconducting proximity 
effect in a ferromagnetic metal \cite{fvl,volkov1,spivak}. In general, one does not
expect to see the proximity effect in a ferromagnet due to the large
internal exchange field which is expected to destroy
superconducting correlations in the ferromagnet at distances greater than the
exchange length $\l_{ex}$ (typically a few nanometers for the
transition metal ferromagnets). This point of view has been reinforced by many
experiments on ferromagnet/superconductor (FS) multilayers, where it was found that
two superconducting layers are effectively decoupled if the thickness of the
ferromagnet between them is much greater than 
$\l_{ex}$ \cite{kawaguchi,chien}.  

More recently, attention has focused on mesoscopic FS
structures, where experimental results seem to indicate that
superconducting correlations can penetrate into the ferromagnet at distances much
greater than $\l_{ex}$. Giroud \etal \cite{giroud} measured the temperature dependent
resistance of mesoscopic Co rings in contact with a superconducting Al film, and
found a small but significant temperature and bias dependent differential
resistance, reminiscent of the reentrant proximity effect observed in normal
metal/superconductor (NS) structures. Petrashov \etal \cite{petrashov1} measured Ni wires in
contact with Al films, and observed an anomalously large change in the resistance
of the devices below the transition temperature of the 
superconductor. This change
was also reflected in the differential resistance of the devices as a function
of dc current below the superconducting transition.

In this Letter, we present results of our measurements of the
resistance of mesoscopic Ni/Al structures as a function of temperature, dc current
and magnetic field. In contrast
to previous experiments, the devices have multiple non-magnetic Au probes which
allow us to separately probe the resistance of different regions of the sample.  In
agreement with previous experiments, we find large changes in resistance below the
superconducting transition of the Al. However, the multiprobe nature of our
devices allows us to determine that the primary contribution to this resistance
change in our samples arises from the FS interface itself, with essentially no contribution from
the ferromagnet, indicating the absence of long range superconducting
correlations in the ferromagnet. In addition, we find that the 
interface resistances of our
devices are sensitive to the magnetic state of the ferromagnetic 
particle. The
resistance of the interface can be reasonably well described by the model of
Blonder, Tinkham and Klapwijk (BTK) \cite{btk}, taking into account the effects of
partial spin polarization of the conduction electrons in the 
ferromagnet \cite{soulen1,soulen2}. 

\epsfig{figure=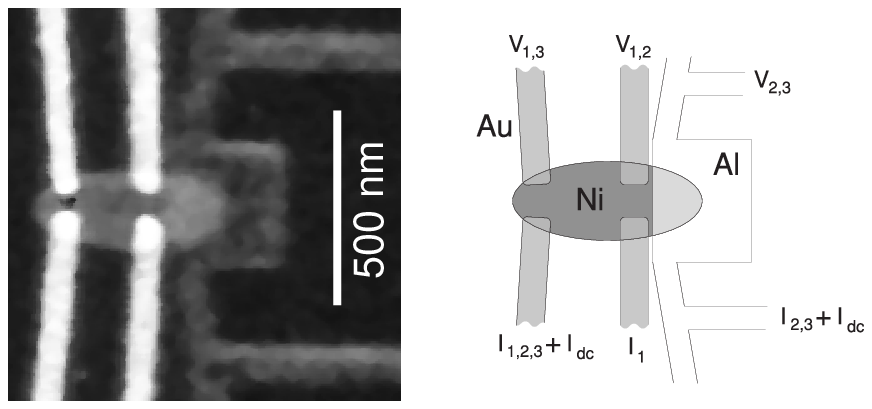}
\begin{figure}
\caption{(a) Micrograph of a typical FS structure.  The picture area is 
scaled to 1 \um $\times$ 1 \um. (b) Schematic of probe configuration. 
The various probe configurations are denoted by the subscripts as 
referred to in the text.}
\label{device}
\end{figure}

Our samples are fabricated in three separate e-beam lithography 
steps with the metals deposited by e-gun deposition. Seven different samples
were measured, but we present here results on only a few 
representative samples. Figures 1(a) and (b) show
a scanning electron micrograph of one of our samples along with a sample
schematic. The majority of our devices consist of an elliptical Ni particle
in contact with a superconducting Al film \cite{foot1}. To ensure predictable
magnetic behavior, the Ni elements are patterned and deposited first so that they
lay flat on  the substrate, and the elliptical shape of the Ni particles ensures
that the magnetic shape anisotropy aligns the magnetization
of the particle in-plane along the major axis of the ellipse \cite{aumentado1}. 
Au wires are then patterned and deposited, contacting the Ni particle and providing
nonmagnetic electronic probes with which we can monitor the magnetic response as well
as measure any proximity effect independent of the response of the FS interface. 
The superconducting layer is then deposited in the final lithography step. All
interfaces are cleaned using an ac Ar$^+$ etch prior to the deposition of the Au and
Al layers. The thickness of the Ni films is $\sim$30 nm, the Al film 
$\sim$50--60 nm,
and the Au electrodes $\sim$50--60 nm. In addition to the FS samples themselves, control
samples of Ni wires, Al wires and Ni/Al interface samples are also fabricated
simultaneously in order to characterize the material parameters of the films and
interfaces. From low temperature measurements on these control samples, the
resistivity of the Ni film was estimated to be $\rho_{Ni}\sim$6.6~\uohmcm and 
that of the Al film $\rho_{Al}\sim$8.4~\uohmcm, corresponding to electronic diffusion constants
$D=(1/3) v_{F} l$ (where $v_{F}$ is the Fermi velocity and $l$ the 
elastic mean free path) of $D_{Ni}\sim$76~cm$^{2}$/s and 
$D_{Al}\sim$26~cm$^{2}$/s respectively \cite{foot2}. 

The measurements are performed at temperatures down to $\sim$260 mK
using standard ac lock-in techniques with all magnetic fields applied \emph{in-plane}
along the easy  axis of the Ni particles using a superconducting split-coil magnet.
The  application of such a longitudinal, in-plane magnetic field is 
advantageous in two respects: first, the critical field of the Al is 
much greater in this configuration, and second, the 
magnetization of the elliptical particles lies in-plane and is single 
domain at remanence \cite{aumentado1}. With this geometry, a number of four-probe measurement
configurations are  possible (see Fig.\ \ref{device}(b)). In this Letter we 
concentrate on only three (as denoted by the subscripts in the figure).  
In configuration ``1'' we measure the resistance of the Ni particle 
while configuration ``2'' measures the interface (with a small 
contribution from the Ni that gaps the distance between the Ni and Al 
probes). Configuration ``3'' measures both the interface and the Ni 
particle resistance in series, and is equivalent to the probe geometry used in 
Ref. \cite{petrashov1}. Measurements which include the interface in
the current path are performed with an excitation current of 10--50 nA, while the Ni particle
measurements are taken with 100--500 nA, low enough to avoid self-heating.

Figure \ref{temp}(a) shows the zero-field temperature dependences
of the  resistances of the FS interface ($R_{2}$) and the FS interface in 
series with the Ni ellipse ($R_{3}$). The normal state resistance of 
the interface in this device was 23.8 $\Omega$. The magnetic state 
of the particle was prepared by saturating the magnetization in a magnetic 
field of +4~kG aligned along the major axis of the elliptical Ni particle 
such that it
contained no domain structure at remanence. The  resistances $R_{2}$ and $R_{3}$
both display a sharp increase at the superconducting transition, and then decrease
until the  temperature reaches 0.9 K, below which the resistances begin to rise 
again. $R_{3}$ simply duplicates that of the interface $R_{2}$, being offset from it by
approximately 2 $\Omega$, which corresponds to the resistance of the Ni 
particle itself. The temperature dependence of 
$R_{3}$ is reminiscent of the reentrant proximity effect seen in 
normal metal mesoscopic structures in contact with 
superconductors \cite{charlat}, and if one had
access to these data alone, one might conclude that the ferromagnet exhibits a strong
superconducting proximity effect. However, a similar resistance change is
\emph{not} seen in the Ni particle by itself (Fig.\ \ref{temp}(b)), indicating that the
resistance change arises in the region of the sample between the voltage probes  of
configuration ``2'', \emph{i.e.}, the FS interface. 
Similar behavior is also observed in our other samples with barrier
resistances ranging from 19 $\Omega$ to 1.3 M$\Omega$. We therefore conclude that no long range
superconducting coherence effects are present in the ferromagnet.

\epsfig{figure=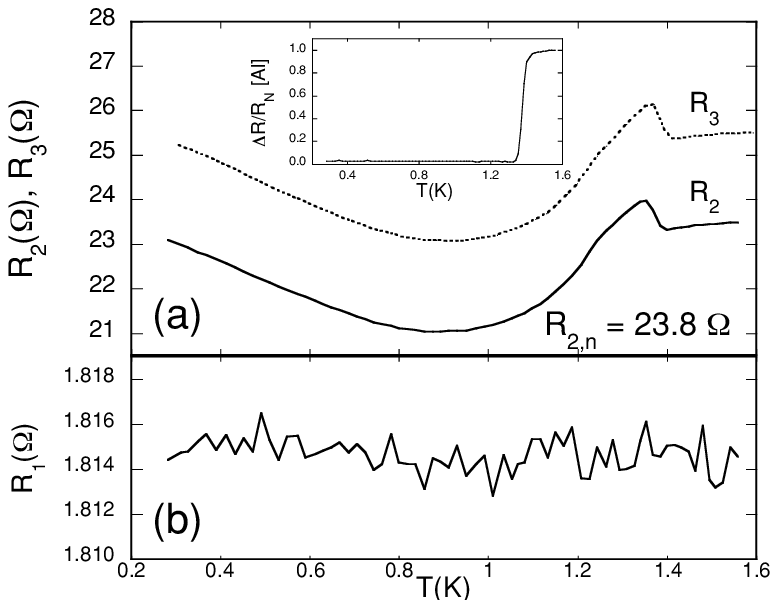}
\begin{figure}
\caption{(a) Temperature dependence of the interface resistance, 
$R_{2}=23.8 \Omega$,  and the interface resistance and Ni ellipse in series, 
$R_{3}$. Inset: the resistance of the overlapping Al wire, $R_{Al}$. 
(b) the resistance of the Ni ellipse, $R_{1}$.}
\label{temp}
\end{figure}

We believe that the peak in the resistance observed near the superconducting transition 
in Fig.\ \ref{temp}(a) is associated with charge imbalance effects in the Al films.  
This can be seen by comparing
the data for low and high interface resistance samples.
Figure~\ref{qpbtk}(a) shows the resistance normalized to the normal state 
value ($r_{2}=R_{2}/R_{2,n}$) of four samples with interface resistances ranging from $\sim$23.8 $\Omega$
to 1.4 M$\Omega$ as a function of temperature in zero applied 
magnetic field. The peak in resistance observed in the low
interface resistance sample disappears as the resistance of the interface increases.  Below
the resistance peak, the data can be reasonably well described by the BTK 
theory with suitable modifications to account for spin polarization 
as we describe below.

\epsfig{figure=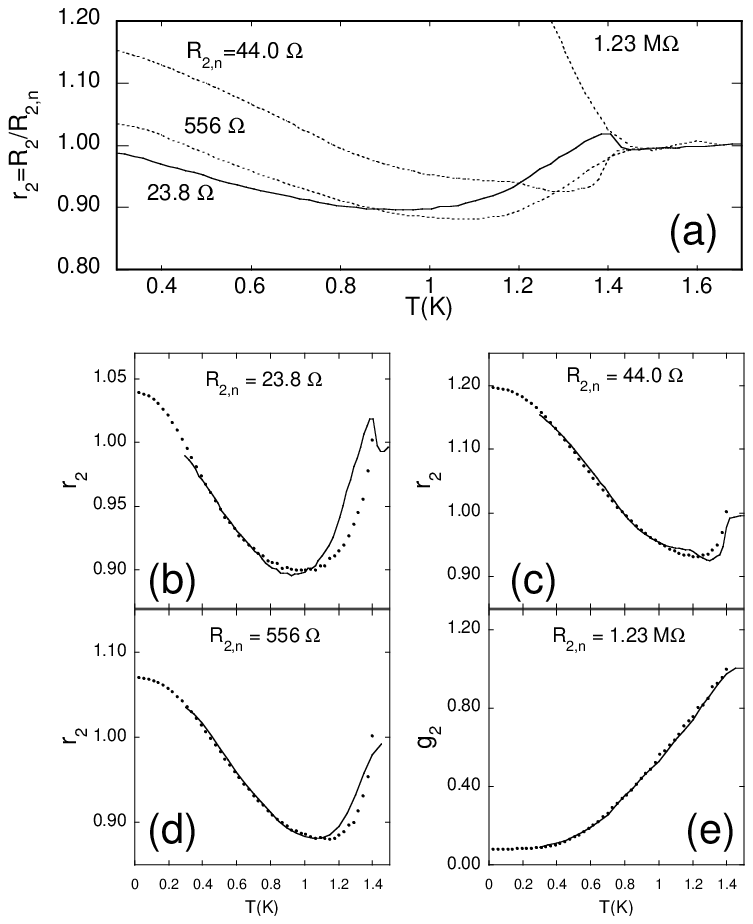}
\begin{figure}
\caption{(a) Normalized temperature dependence for various values of the normal 
state barrier resistance, $R_{2,n}$ (as noted in figure). Solid trace 
(lowest barrier resistance) shows a charge imbalance peak near 
$T_{c}$(=1.4 K). (b)--(e) BTK fits (modified to include the effect of spin 
polarization, $P$) for various values of the interface resistance, 
$R_{2,n}$. See text for fitting parameters. ((b)--(d) Normalized 
resistances, (e) normalized conductance.)}
\label{qpbtk}
\end{figure}

The normalized conductance of an NS point contact in the BTK model 
is \cite{btk}
\begin{equation}
g(Z,T) = (1+Z^{2})\int_{-\infty}^{+\infty}\left(-\frac{\partial f_{0}}{\partial E}\right)[1+A(E)-B(E)]dE,
\label{eq:eqn1}  
\end{equation}
where $f_{0}$ is the Fermi function, and $A(E)$ and $B(E)$ are the BTK
parameters which describe Andreev and normal reflection processes respectively. $A(E)$ and
$B(E)$ depend on the gap in the superconductor $\Delta$ and the BTK parameter $Z$ which
parameterizes the strength of the interface.  In the case when the normal 
metal is a ferromagnet (FS transport), the spin-polarization 
$P=(N_{\uparrow}(E_{F})-N_{\downarrow}(E_{F}))/(N_{\uparrow}(E_{F})+N_{\downarrow}(E_{F}))$ of the 
electrons in the ferromagnet must be considered. Since Andreev reflection 
processes can 
only occur between pairs of spin-up and spin-down electrons, the 
fraction of the electrons that can participate in such a process is 
$(1-P)$ of the total population. To account for this in the BTK 
model \cite{btk}, one may replace the factor $A(E)$ in equation (\ref{eq:eqn1}) with 
$A'(E) = (1-P) A(E)$ \cite{soulen1,soulen2}. This substitution was performed 
by Soulen \etal \cite{soulen1,soulen2} to determine the polarization of various ferromagnetic metals using 
point contact spectroscopy in clean contacts. Using this same substitution, one may fit the 
temperature dependence for arbitrary values of $Z$ and $P$. 

The dotted traces in Fig.\ \ref{qpbtk}(b)--(e) show numerical
fits of our data (solid traces) to the normalized resistance (or conductance) predicted by 
the modified BTK theory for different 
values of $Z$, $P$. In our model we also allow for magnetic flux 
penetration into the superconductor from the field generated by the 
ferromagnet near the interface. This necessitates another free 
parameter in the fitting routine since it is difficult to predict the 
exact flux penetration profile near the interface. We found that fixing $P$ at zero nearly 
always gave inferior fits to those performed with $P$ as a free 
parameter. For the traces shown in Fig.\ \ref{qpbtk}(b)--(d), the $Z$ values were all 
similar (0.38$<Z<$0.50), while the best fits where found with 
0.21$<P<$0.30, in rough agreement with the value, $P_{Ni}\sim$ 0.23 
found by FS tunnelling spectroscopy \cite{tmreview}. Our highest 
resistance sample (Fig. \ref{qpbtk}(e)) fit with a higher value of 
$Z=2.1$, while also yielding a polarization $P=0.28$. We also observe evidence for a 
finite spin polarization in the differential 
resistance as a function of dc current, although these data are not 
discussed here.

In contrast to previous FS experiments, in many of our devices
two or more distinct states were seen in the 
temperature dependence of the interface(see Fig.\ \ref{field}(a)); the 
samples frequently showed switching between these states
while the sample temperature was swept. These multiple 
states were also seen in the magnetic field dependence of the 
interface at fixed temperature. Figure \ref{field}(b) shows a number 
of magnetoresistance (MR) traces for both the interface ($R_{2}$) and the 
overlapping Al ($R_{Al}$), with field sweeps in both positive and negative directions. 
There is a strong low-field dependence with sharp jumps at $+350$ G 
and $-300$ G. A MR trace of the Ni ellipse by itself 
shows standard AMR behavior (see Ref. \cite{aumentado1}) with sharp 
jumps at exactly the same fields (see Fig.\ \ref{field}(c)). Since these jumps are due to the 
switching of the magnetization from positive to negative orientation 
(and vice versa), it is clear that the interface resistance, $R_{2}$, 
is sensitive to the local field generated by the ferromagnet itself. 
Even with no applied field, the ferromagnet may generate a substantial 
amount of flux and should never be assumed to vanish, especially in 
this geometry. Furthermore, the absence of 
multiple states in the Ni MR (for positive or negative magnetization orientation) 
suggests that the states seen in the temperature and field 
dependences of the interface are due to multiple magnetic screening 
states in the superconductor itself.

\epsfig{figure=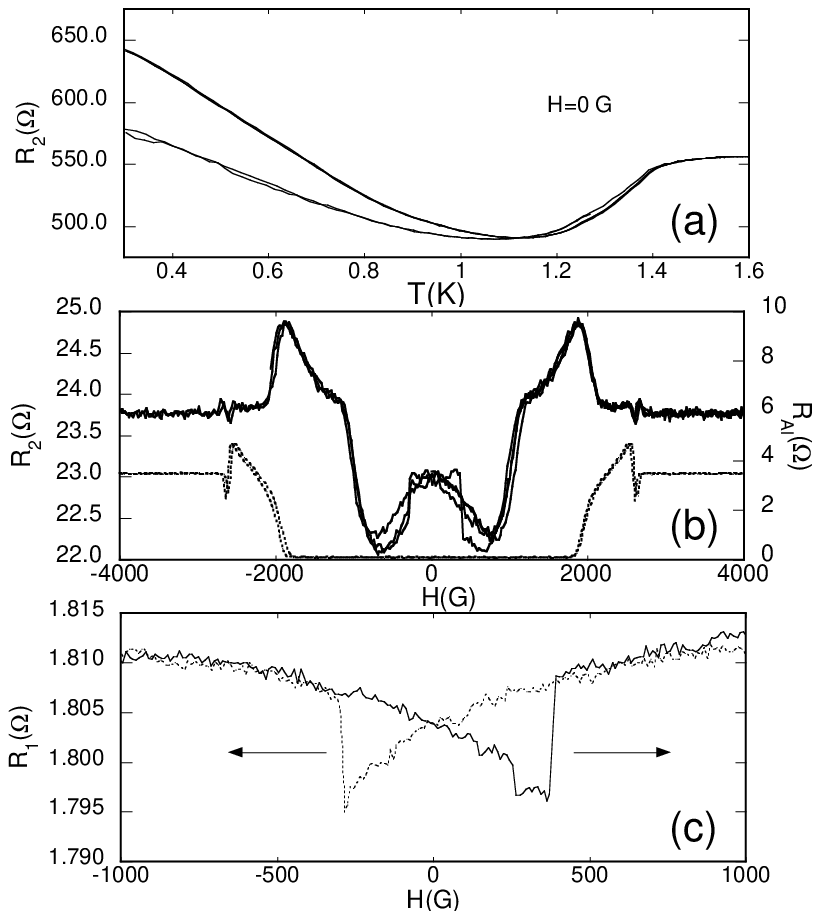}
\begin{figure}
\caption{(a) Multiple states in the temperature dependence of the 556~$\Omega$ 
interface resistance sample; MRs (at $T=$300 mK): (b) interface 
resistance, $R_{2}$, showing 
multiple states(left axis, solid trace). 
overlapping Al wire, $R_{Al}$ (right axis, dashed trace). (c) Ni 
ellipse, $R_{1}$ (arrows indicate sweep direction).}
\label{field}
\end{figure}

Although we have restricted the above analysis to a 
modified BTK model, recent work by Golubov \cite{golubov} has modified 
the BTK model to account for charge-imbalance and diffusive 
interfaces, while Belzig 
\etal \cite{belzig} have analyzed dirty and diffusive FS interfaces 
within the framework of nonequilibrium Green's function theory. 
While these approaches are certainly more sophisticated than our 
simple approach, qualitatively they predict behavior similar to our 
experimental results for the temperature dependence. However, to our 
knowledge there is no available published work which includes 
charge-imbalance, spin-accumulation, and the effect of field 
penetration into the superconductor. In addition to these effects, a 
complete theory should include effects of spin-splitting in $N_{s}(E)$, 
since even at zero applied 
magnetic field, the superconductor may be subjected to a substantial magnetic field 
generated by the ferromagnet very close to the interface. This is 
further complicated by the fact that such a field may not be 
homogeneous with respect to the superconducting coherence length. 
Although we have attempted to establish as uniform a field 
distribution as possible (by carefully selecting an elliptical 
geometry), ultimately it is very difficult to construct a device in 
which the field penetration in the superconductor is uniform at 
$H_{applied}\ll H_{c}$. Furthermore, at finite voltage or current bias
it is possible that the charge-imbalance is strongly modified by the 
spin-polarized quasiparticle current that is generated at the interface. 
Since these quasiparticle excitations are expected to decay into 
Cooper pairs, matching spin-up and spin-down electrons equally, the 
spin-imbalance may prolong the quasiparticle decay time 
$\tau_{Q^{*}}$ substantially if the 
spin-scattering lifetime $\tau_{s}$ is much larger.
In essence, a complete theory of FS transport will need to include 
the nonequilibrium superconductivity, spin-accumulation in both F and 
S, and spin-splitting of $N_{s}$. 

In summary, our results are in agreement with recent theoretical work 
which suggest that a proximity effect within the ferromagnet is 
negligible, while the main contribution to the resistance change is due 
to the interface. However, the effects of finite field and 
charge-imbalance may be important in constructing a comprehensive 
theory of FS transport in mesoscopic structures.

This work was supported by the David and Lucile
Packard Foundation and the National Science Foundation
through DMR-9801982.

%
%
%

\end{multicols}


\begin{thebibliography}{10}
\bibitem{foot1}
We have also measured circular and rectangular elements in similar
  configurations. Although these results are essentially similar to the
  elliptical element results they suffer from nonuniform magnetization
  distributions in small magnetic fields. For this reason, we concentrate only
  on elliptical magnets in this Letter.
\end{thebibliography}
\end{document}